\documentclass[12pt]{article}

\usepackage{amssymb}
\makeatletter \@addtoreset{equation}{section} \makeatother

\def\ftoday{{\sl {Le \number\day \space\ifcase\month
\or janvier\or f\'evrier\or mars\or avril\or mai \or juin\or
juillet\or ao\^ut\or septembre\or octobre \or novembre \or
d\'ecembre\fi\space \number\year}}}
\def\ptoday{{\sl {\number\day \space de\space \ifcase\month
\or janeiro\or fevereiro\or mar{\c c}o\or abril\or maio \or
junho\or julho\or agosto\or setembro\or outubro \or novembro \or
dezembro\fi\space de\space \number\year}}}
\def\gtoday{{\sl {Den \number\day. \ifcase\month
\or Januar\or Februar\or M\"arz\or April\or Mai \or Juni\or
Juli\or August\or September\or Oktober \or November \or
Dezember\fi\space \number\year}}}
\def\today{{\sl {\ifcase\month
\or January\or February\or March\or April\or May \or June\or
July\or August\or September\or October \or November \or
December\fi \space\number\day,\space
                                            \number\year}}}

\newcommand{\XI}{\XI}

\newcommand{\sla}{\raise.15ex\hbox{$/$}\kern -.57em}
\newcommand{\Sla}{\raise.15ex\hbox{$/$}\kern -.70em}

\newcommand{\complex}{{\kern .1em {\raise .47ex
\hbox {$\scriptscriptstyle |$}}
    \kern -.4em {\rm C}}}
\newcommand{\real}{{{\rm I} \kern -.19em {\rm R}}}
\newcommand{\rational}{{\kern .1em {\raise .47ex
\hbox{$\scripscriptstyle |$}}
    \kern -.35em {\rm Q}}}
\renewcommand{\natural}{{\vrule height 1.6ex width
.05em depth 0ex \kern -.35em {\rm N}}}

\newcommand{\twiddle}{\lower.9ex\rlap{$\kern -.1em\scriptstyle\sim$}}

\newcommand{\eq}{\begin{equation}}
\newcommand{\eqn}[1]{\label{#1}\end{equation}}
\newcommand{\eea}{\end{eqnarray}}
\newcommand{\eqa}{\begin{eqnarray}}
\newcommand{\eqan}[1]{\label{#1}\end{eqnarray}}
\newcommand{\ba}{\begin{array}}
\newcommand{\ea}{\end{array}}
\newcommand{\eqac}{\begin{equation}\begin{array}{rcl}}
\newcommand{\eqacn}[1]{\end{array}\label{#1}\end{equation}}

\begin{document}



\begin{center}
{\Large \bf Recent Results on $N=2,4$ Supersymmetry with Lorentz
Symmetry Violating}
\end{center}
\vspace{3mm}

\begin{center}{\large
Wander G. Ney$^{a,b,c}$, J. A. Helayel-Neto$^{b,c}$ and Wesley
Spalenza$^{b,c,d,}$\footnote{Supported by the Conselho Nacional de
Desenvolvimento Cient\'{\i}fico e Tecnol\'{o}gico CNPq -
Brazil.\\
{\tt E-mails: wander@cbpf.br, helayel@cbpf.br, spalenza@sissa.it}}
} \vspace{1mm}

\noindent
$^{a}$Centro Federal de Educa\c{c}\~{a}o Tecnol\'{o}gica de Campos (CEFET) RJ-Brazil\\
$^{b}$Centro Brasileiro de Pesquisas F\'{i}sicas (CBPF) RJ-Brazil\\
$^{c}$Grupo de F\'{i}sica Te\'{o}rica Jos\'{e} Leite Lopes (GFT)\\
$^{d}$Scuola Internazionale Superiore di Studi Avanzati (SISSA)
Trieste-Italy.
\end{center}
\vspace{1mm}

{\hfill\parbox{198mm}{{\textbf{Fourth International Winter
Conference on \\ Mathematical Methods in Physics} \\
\\
\textit{Centro Brasileiro de Pesquisas F\'{i}sicas (CBPF/MCT)\\
 Rio de Janeiro, Brazil \\
 09 - 13 August 2004}}}} \vspace{3mm}


\abstract{In this work, we propose the $N=2$ and $N=4$
supersymmetric extensions of the Lorentz-breaking Abelian
Chern-Simons term. We formulate the question of the Lorentz
violation in $6$ and $10$ dimensions to obtain  the bosonic
sectors of $N=2-$ and $N=4-$ supersymmetries, respectively. From
this, we carry out an analysis in $N=1-\,D=4$ superspace and, in terms of $%
N=1-$ superfields, we are able to write down the $N=2$ and $N=4$
supersymmetric extensions of the Lorentz-violating action term.}

\section{Introduction}

The formulation of physical models for the fundamental
interactions in the framework of quantum field theories for
point-like objects is based on a number of principles, among which
Lorentz covariance and invariance under suitable gauge symmetries.
However, mechanisms for the breakdown of these symmetries have
been proposed and discussed in view of a number of
phenomenological and experimental evidences \cite{1,1b,1c,1d,1e}.
Astrophysical observations indicate that Lorentz symmetry may be
slightly violated in order to account for anisotropies. Then, one
may consider a gauge theory where Lorentz symmetry breaking may be
realized by means of a term in the action. A Chern-Simons-type
term may be considered that exhibits a constant background
four-vector which maintains the gauge invariance but breaks down
the Lorentz space-time symmetry \cite{1}.

In the context of supersymmetry (SUSY), the issue of Lorentz
violation has been considered in the literature in different
formulations: in ref. \cite{3a}, supersymmetry is presented by
introducing a suitable modification in its algebra; in ref.
\cite{3,3c}, one achieves the $N=1-$SUSY version of the
Chern-Simons term by means of the conventional
superspace-superfield formalism; in ref. \cite{3b}, the authors
adopt the idea of Lorentz breaking operators. More particularly,
considering the importance of extended supersymmetries in
connection with gauge theories, we propose in this work an $N=2$
and an $N=4$ extended supersymmetric generalization of the
Lorentz-breaking Chern-Simons term in a
4-dimensional Minkowski background. We start off with the Chern-Simons term in $%
(1+5)$ and $\left( 1+9\right) $ space-time dimensions and adopt a
particular
dimensional reduction method, see \cite{2}, to obtain the bosonic sector in $%
D=(1+3)$ of the $N=2$ and $N=4$ supersymmetric models,
respectively. This is possible because in $N=1,D=6$- and
$N=1,D=10$-supersymmetries, the bosonic sector has the same number
of degrees of freedom as the bosonic sector of an $N=2,D=4$ and
$N=4,D=4$, respectively \cite{2b}. Once the bosonic sectors are
identified, we adopt an $N=1,D=4$-superfield formulation to write
down the gauge potential and the Lorentz-violating background
supermultiplets to finally set up their coupling in terms of $N=2$
and $N=4$ actions realized in $N=1$-superspace. The result is
projected out in component fields and we end up with the complete
actions that realize the extended supersymmetric version of the
Abelian Chern-Simons Lorentz-violating term.

\section{$N=2$-Lorentz-violating term}

The $N=2$ supersymmetric generalization of the Abelian
Chern-Simons Lorentz breaking term can be built up using
superfield formalism in an $N=1$ superspace background, having
coordinates $(x^{\mu },\theta ^{a},\bar{\theta}_{\dot{a}})$
\cite{2}. Using the fact that the bosonic sector for $N=2$
in $
D=6$ and make the dimensional reduction for $D=4$. The $D=4$
Chern-Simons term proposed originally by \cite{1} is
\begin{equation}
{\cal L}_{br}=\varepsilon ^{\mu \nu \kappa \lambda }A_{\mu
}\partial _{\nu }A_{\kappa }T_{\lambda }.  \label{7}
\end{equation}
We propose for $D=6$ the Chern-Simons term in the form
\begin{equation}
{\cal L}_{br}=\varepsilon ^{\hat{\mu}\hat{\nu}\hat{\kappa}\hat{\lambda}\hat{%
\rho}\hat{\sigma}}A_{\hat{\mu}}\partial _{\hat{\nu}}A_{\hat{\kappa}}T_{\hat{%
\lambda}\hat{\rho}\hat{\sigma}},  \label{7a}
\end{equation}
where $\hat{\mu}=\mu ,4,5$. The gauge field has 6 components, then
we redefine it as $A_{\hat{\mu}}\equiv (A_{\mu };\varphi
_{1};\varphi _{2}).$ The background tensor
$T_{\hat{\lambda}\hat{\rho}\hat{\sigma}}$ has 20 components, but
we can redefine it as
$T_{\hat{\lambda}\hat{\rho}\hat{\sigma}}\equiv (R_{\rho \sigma
};S_{\rho \sigma };\partial _{\mu }v;\partial _{\mu }u)$.
The fields $R_{\rho \sigma }$ and $%
S_{\rho \sigma }$ has 6 components each one, and the other 8
components are redefined as 2 vectors that we write as a gradient
of the scalars fields $v$ and $u$. Then, the number of components
is reduced to 14. The dimensional reduction is done considering
that there are not dependence of the fields in the $x^{4},x^{5}$
coordinates. It is clear that $\varepsilon
^{\hat{\mu}\hat{\nu}\hat{\kappa}\hat{\lambda}\hat{\rho}\hat{\sigma}}A_{\hat{%
\mu}}A_{\hat{\kappa}}\partial _{\hat{\nu}}T_{\hat{\lambda}\hat{\rho}\hat{%
\sigma}}=0,$ so we obtain, integrating by parts, the reduced
Lagrangian as follows:
\begin{eqnarray}
{\cal L}_{br} &=&-\frac{1}{4}\varepsilon ^{\mu \nu \kappa \lambda
}F_{\mu \nu }A_{\kappa }\partial _{\lambda
}v+\frac{1}{4}\varepsilon ^{\mu \nu
\kappa \lambda }F_{\mu \nu }\varphi _{1}R_{\kappa \lambda }+\frac{1}{4}%
\varepsilon ^{\mu \nu \kappa \lambda }F_{\mu \nu }\varphi
_{2}S_{\kappa
\lambda }  \label{8} \\
&&+\frac{1}{2}\varphi _{1}\partial _{\nu }\varphi _{2}\partial ^{\nu }u-%
\frac{1}{2}\varphi _{2}\partial _{\nu }\varphi _{1}\partial ^{\nu
}u. \nonumber
\end{eqnarray}
In order to make the supersymmetrization of the Lagrangian
(\ref{8}) using the superspace formalism, we have to define some
complex fields that can be found in superfields. We define these
bosonic fields as
\begin{eqnarray}
B_{\mu \nu } &=&S_{\mu \nu }-i\tilde{S}_{\mu \nu },  \nonumber \\
H_{\mu \nu } &=&R_{\mu \nu }-i\tilde{R}_{\mu \nu }  \nonumber \\
\varphi &=&\varphi _{1}+i\varphi _{2},  \label{8b} \\
r &=&t+iu,  \nonumber \\
s &=&w+iv,  \nonumber
\end{eqnarray}

Notice that we have introduced the new real scalar fields $t$ and
$w$ that are bosonic fields that do not appear in the bosonic
Lagrangian (\ref{8}). These fields will be necessary in the
supersymmetric version to maintain the same number of degree of
freedom between bosonic and fermionic sector due the scalar
superfields are defined with complex scalar fields. Each tensor
field, $R_{\mu \nu }$ and $S_{\mu \nu },$ appears as the real part
of the complex tensor field whose imaginary parts are given in
terms of their dual fields, as we see in (\ref{8b}) and can be
found in \cite{5}. The vector superfield $V$ that accommodates
$A_{\mu}$ in the WZ-gauge is written as:
\begin{equation}
V=\theta \sigma ^{\mu }\bar{\theta}A_{\mu }+\theta ^{2}\bar{\theta}\bar{%
\lambda}+\bar{\theta}^{2}\theta \lambda +\theta
^{2}\bar{\theta}^{2}D, \label{3}
\end{equation}
which fulfills the reality constraint, $V=V^{\dagger }.$\ The
scalar superfield that accommodates $\varphi$ and $\varphi ^{*}$
is written as
\begin{equation}
\Phi =\varphi +i\theta \sigma ^{\mu }\bar{\theta}\partial _{\mu }\varphi -%
\frac{1}{4}\theta ^{2}\bar{\theta}^{2}\Box \varphi +\sqrt{2}\theta \psi +%
\frac{i}{\sqrt{2}}\theta ^{2}\partial _{\mu }\psi \sigma ^{\mu }\bar{\theta}%
+\theta ^{2}f,  \label{4}
\end{equation}
and this complex conjugate $\bar{\Phi}$. These superfields obey
the chiral condition: $\bar{D}\Phi =D\bar{\Phi}=0.$\ The scalar
superfields that accommodate $s,\,r$ and $r$ and their respective
complex conjugate fields are:
\begin{equation}
S=s+i\theta \sigma ^{\mu }\bar{\theta}\partial _{\mu
}s-\frac{1}{4}\theta ^{2}\bar{\theta}^{2}\Box s+\sqrt{2}\theta \xi
+\frac{i}{\sqrt{2}}\theta ^{2}\partial _{\mu }\xi \sigma ^{\mu
}\bar{\theta}+\theta ^{2}h,  \label{10}
\end{equation}
\begin{equation}
R=r+i\theta \sigma ^{\mu }\bar{\theta}\partial _{\mu
}r-\frac{1}{4}\theta ^{2}\bar{\theta}^{2}\Box r+\sqrt{2}\theta
\zeta +\frac{i}{\sqrt{2}}\theta ^{2}\partial _{\mu }\zeta \sigma
^{\mu }\bar{\theta}+\theta ^{2}g, \label{11}
\end{equation}
and their complex conjugate superfields $\bar{S}$ and $\bar{R}$
which satisfy the chiral condition:
$\bar{D}S=D\bar{S}=\bar{D}R=D\bar{R}=0.$
 The spinor superfields
that contain $R_{\mu \nu },S_{\mu \nu }$ and their corresponding
dual fields are written as
\begin{eqnarray}
\Sigma _{a} &=&\tau _{a}+\theta ^{b}(\varepsilon _{ba}\rho +\sigma
_{ba}^{\mu \nu }B_{\mu \nu })+\theta ^{2}F_{a}+i\theta \sigma ^{\mu }\bar{%
\theta}\partial _{\mu }\tau _{a}  \label{12} \\
&&+i\theta \sigma ^{\mu }\bar{\theta}\theta ^{b}\partial _{\mu
}(\varepsilon
_{ba}\rho +\sigma _{ba}^{\mu \nu }B_{\mu \nu })-\frac{1}{4}\theta ^{2}\bar{%
\theta}^{2}\square \tau _{a},  \nonumber
\end{eqnarray}
\begin{eqnarray}
\Omega _{a} &=&\chi _{a}+\theta ^{b}(\varepsilon _{ba}\phi +\sigma
_{ba}^{\mu \nu }H_{\mu \nu })+\theta ^{2}G_{a}+i\theta \sigma ^{\mu }\bar{%
\theta}\partial _{\mu }\chi _{a}  \label{12a} \\
&&+i\theta \sigma ^{\mu }\bar{\theta}\theta ^{b}\partial _{\mu
}(\varepsilon
_{ba}\phi +\sigma _{ba}^{\mu \nu }H_{\mu \nu })-\frac{1}{4}\theta ^{2}\bar{%
\theta}^{2}\square \chi _{a},  \nonumber
\end{eqnarray}
and their complex conjugate superfields $\bar{\Sigma}$ and
$\bar{\Omega}$ that are also chiral: $\bar{D}_{\dot{b}}\Sigma
_{a}= D_{b}\bar{\Sigma}_{\dot{a} }=\bar{D}_{\dot{b}}\Omega
_{a}=D_{b}\bar{\Omega}_{\dot{a}}=0.$ We can notice that we have to
introduce two extra background complex scalar fields, $\rho $ and
$\phi ,$ to match the bosonic and fermionic degrees of freedom.

Now, we are interested in building up the supersymmetric action.
For that, we take into consideration the canonical (mass)
dimensions of the superfields; based on these dimensionalities,
and by analyzing the bosonic Lagrangian (\ref {8}), we propose the
following supersymmetric action, $S_{br}$:

\begin{eqnarray}
{\cal S}_{br} &=&\int d^{4}xd^{2}\theta d^{2}\bar{\theta}[\frac{1}{4}%
W^{a}(D_{a}V)S+\frac{1}{4}\bar{W}_{\dot{a}}(\bar{D}^{\dot{a}}V)\bar{S}+\frac{%
i}{4}\delta (\bar{\theta})W^{a}(\Phi +\bar{\Phi})\Sigma _{a}  \nonumber \\
&&-\frac{i}{4}\delta (\theta )\bar{W}_{\dot{a}}(\Phi +\bar{\Phi})\bar{\Sigma}%
^{\dot{a}}+\frac{1}{4}\delta (\bar{\theta})W^{a}(\Phi
-\bar{\Phi})\Omega _{a}
\label{13} \\
&&-\frac{1}{4}\delta (\theta )\bar{W}_{\dot{a}}(\Phi -\bar{\Phi})\bar{\Omega}%
^{\dot{a}}+\frac{1}{4}\Phi \bar{\Phi}(\bar{R}+R)],  \nonumber
\end{eqnarray}

We therefore observe that the action (\ref{13}) is manifestly
invariant under $N=1$-supersymmetry. The component-field content
of the $N=2$-supersymmetry is accommodated in the
$N=1$-superfields). Indeed, the action (\ref{13}) displays a
larger supersymmetry, $N=2,$ realized in terms of an $N=1$
-superspace formulation.

This Lagrangian in its component-field version reads as below:
\begin{eqnarray}
{\cal L}_{br} &=&+\frac{i}{8}\partial _{\mu }(s-s^{*})\varepsilon
^{\mu \kappa \lambda \nu }F_{\kappa \lambda }A_{\nu
}-\frac{1}{8}(s+s^{*})F_{\mu
\nu }F^{\mu \nu }+D^{2}(s+s^{*})  \nonumber \\
&&-\frac{1}{2}is\lambda \sigma ^{\mu }\partial _{\mu }\bar{\lambda}-\frac{1}{%
2}is^{*}\bar{\lambda}\bar{\sigma}^{\mu }\partial _{\mu }\lambda -\frac{1}{2%
\sqrt{2}}\lambda \sigma ^{\mu \nu }F_{\mu \nu }\xi +\frac{1}{2\sqrt{2}}\bar{%
\lambda}\bar{\sigma}^{\mu \nu }F_{\mu \nu }\bar{\xi}  \nonumber \\
&&+\frac{1}{4}\lambda \lambda h+\frac{1}{4}\bar{\lambda}\bar{\lambda}h^{*}-%
\frac{1}{\sqrt{2}}\lambda \xi
D-\frac{1}{\sqrt{2}}\bar{\lambda}\bar{\xi}D
\nonumber \\
&&\frac{1}{16}\varepsilon ^{\mu \nu \kappa \lambda }F_{\mu \nu
}(\varphi
+\varphi ^{*})(B_{\kappa \lambda }+B_{\kappa \lambda }^{*})+\frac{i}{8}%
F^{\mu \nu }(B_{\mu \nu }-B_{\mu \nu }^{*})(\varphi +\varphi ^{*})
\nonumber
\\
&&-\frac{i\sqrt{2}}{8}\tau \sigma ^{\mu \nu }\psi F_{\mu \nu }-\frac{i\sqrt{2%
}}{8}\bar{\tau}\bar{\sigma}^{\mu \nu }\bar{\psi}F_{\mu \nu
}+\frac{1}{4}\tau \sigma ^{\mu }\partial _{\mu
}\bar{\lambda}(\varphi +\varphi ^{*})  \nonumber
\\
&&-\frac{1}{4}\bar{\tau}\bar{\sigma}^{\mu }\partial _{\mu }\lambda
(\varphi +\varphi ^{*})+\frac{i\sqrt{2}}{4}\psi \sigma ^{\mu \nu
}B_{\mu \nu }\lambda
+\frac{i\sqrt{2}}{4}\bar{\psi}\bar{\sigma}^{\mu \nu }B_{\mu \nu }^{*}\bar{%
\lambda}  \nonumber \\
&&-\frac{i}{2}D(\varphi +\varphi ^{*})\rho
+\frac{i}{2}D^{*}(\varphi
+\varphi ^{*})\rho ^{*}  \nonumber \\
&&+\frac{i\sqrt{2}}{8}\lambda \psi \rho -\frac{i\sqrt{2}}{8}\bar{\lambda}%
\bar{\psi}\rho ^{*}-\frac{i\sqrt{2}}{4}D\psi \tau +\frac{i\sqrt{2}}{4}D^{*}%
\bar{\psi}\bar{\tau}  \label{14} \\
&&+\frac{i}{4}f\lambda \tau -\frac{i}{4}f^{*}\bar{\lambda}\bar{\tau}+\frac{i%
}{4}(\varphi +\varphi ^{*})\lambda F-\frac{i}{4}(\varphi +\varphi ^{*})\bar{%
\lambda}\bar{F}  \nonumber \\
&&-\frac{i}{16}\varepsilon ^{\mu \nu \kappa \lambda }F_{\mu \nu
}(\varphi
-\varphi ^{*})(H_{\kappa \lambda }+H_{\kappa \lambda }^{*})+\frac{1}{8}%
F^{\mu \nu }(H_{\kappa \lambda }-H_{\kappa \lambda }^{*})(\varphi
-\varphi
^{*})  \nonumber \\
&&-\frac{\sqrt{2}}{8}\chi \sigma ^{\mu \nu }\psi F_{\mu \nu }+\frac{\sqrt{2}%
}{8}\bar{\chi}\bar{\sigma}^{\mu \nu }\bar{\psi}F_{\mu \nu
}-\frac{i}{4}\chi \sigma ^{\mu }\partial _{\mu
}\bar{\lambda}(\varphi -\varphi ^{*})  \nonumber
\\
&&+\frac{i}{4}\bar{\chi}\bar{\sigma}^{\mu }\partial _{\mu }\lambda
(\varphi
-\varphi ^{*})+\frac{\sqrt{2}}{4}\psi \sigma ^{\mu \nu }H_{\mu \nu }\lambda -%
\frac{\sqrt{2}}{4}\bar{\psi}\bar{\sigma}^{\mu \nu }H_{\mu \nu }^{*}\bar{%
\lambda}  \nonumber \\
&&-\frac{1}{2}D(\varphi -\varphi ^{*})\phi
+\frac{1}{2}D^{*}(\varphi
-\varphi ^{*})\phi ^{*}  \nonumber \\
&&+\frac{\sqrt{2}}{8}\lambda \psi \phi +\frac{\sqrt{2}}{8}\bar{\lambda}\bar{%
\psi}\phi ^{*}-\frac{\sqrt{2}}{4}D\psi \chi -\frac{\sqrt{2}}{4}D^{*}\bar{\psi%
}\bar{\chi}  \nonumber \\
&&+\frac{1}{4}f\lambda \chi +\frac{1}{4}f^{*}\bar{\lambda}\bar{\chi}+\frac{1%
}{4}(\varphi -\varphi ^{*})\lambda G-\frac{1}{4}(\varphi -\varphi ^{*})\bar{%
\lambda}\bar{G}  \nonumber \\
&&+\frac{1}{8}\varphi \partial _{\mu }\varphi ^{*}\partial ^{\mu }(r-r^{*})-%
\frac{1}{8}\varphi ^{*}\partial _{\mu }\varphi \partial ^{\mu
}(r-r^{*})
\nonumber \\
&&+\frac{1}{4}\partial ^{\mu }\varphi \partial _{\mu }\varphi ^{*}(r+r^{*})-%
\frac{1}{8}\varphi \varphi ^{*}\square (r+r^{*})-\frac{i}{4}\psi
\sigma
^{\mu }\partial _{\mu }\bar{\psi}(r+r^{*})  \nonumber \\
&&+\frac{1}{4}ff^{*}(r+r^{*})-\frac{i}{4}\psi \sigma ^{\mu }\bar{\psi}%
\partial _{\mu }r^{*}  \nonumber \\
&&-\frac{i}{4}\varphi \zeta \sigma ^{\mu }\partial _{\mu }\bar{\psi}-\frac{i%
}{4}\varphi ^{*}\psi \sigma ^{\mu }\partial _{\mu }\bar{\zeta}-\frac{i}{4}%
\psi \sigma ^{\mu }\bar{\zeta}\partial _{\mu }\varphi ^{*}  \nonumber \\
&&+\frac{1}{4}\varphi f^{*}g+\frac{1}{4}f\varphi ^{*}g^{*}-\frac{1}{4}%
f^{*}\psi \zeta -\frac{1}{4}f\bar{\psi}\bar{\zeta}.  \nonumber
\end{eqnarray}
We point out the pieces corresponding to the bosonic action
(\ref{8}) in the complete component-field action above:
\begin{eqnarray*}
\frac{i}{8}\partial _{\mu }(s-s^{*})\varepsilon ^{\mu \kappa
\lambda \nu }F_{\kappa \lambda }A_{\nu }
&=&-\frac{1}{4}\varepsilon ^{\mu \nu \kappa
\lambda }F_{\mu \nu }A_{\kappa }\partial _{\lambda }v, \\
\frac{1}{16}\varepsilon ^{\mu \nu \kappa \lambda }F_{\mu \nu
}(\varphi
+\varphi ^{*})(B_{\kappa \lambda }+B_{\kappa \lambda }^{*}) &=&\frac{1}{4}%
\varepsilon ^{\mu \nu \kappa \lambda }F_{\mu \nu }\varphi
_{1}R_{\kappa
\lambda }, \\
-\frac{i}{16}\varepsilon ^{\mu \nu \kappa \lambda }F_{\mu \nu
}(\varphi
-\varphi ^{*})(H_{\kappa \lambda }+H_{\kappa \lambda }^{*}) &=&\frac{1}{4}%
\varepsilon ^{\mu \nu \kappa \lambda }F_{\mu \nu }\varphi
_{2}S_{\kappa
\lambda }, \\
\frac{1}{8}\varphi \partial _{\mu }\varphi ^{*}\partial ^{\mu }(r-r^{*})-%
\frac{1}{8}\varphi ^{*}\partial _{\mu }\varphi \partial ^{\mu }(r-r^{*}) &=&%
\frac{1}{2}\varphi _{1}\partial _{\nu }\varphi _{2}\partial ^{\nu }u-\frac{1%
}{2}\varphi _{2}\partial _{\nu }\varphi _{1}\partial ^{\nu }u.
\end{eqnarray*}
We can notice that this Lagrangian describes the bosonic sector
(\ref{8}) and its superpartners. We find here the $N=1$
supersymmetrization of the Chern-Simons term presented in
\cite{3}, where the first term is the same as proposed by
\cite{1}, considering the constant vector as the gradient of a
scalar. Since the gradient vector is a constant, we have that
$s=\alpha +\beta ^{\mu }x_{\mu }.$ We see in the Lagrangian the
presence of the bosonic real scalar fields, $t=s+s^{*}$ and
$u=r+r^{*},$ and the complex scalar fields, $\rho $ and $\phi ,$
that do not appear in the bosonic Lagrangian (\ref{8}). These
scalar fields appear in the supersymmetric generalization in order
to keep the bosonic and fermionic degrees of freedom in equal
number. We point out  that the bosonic fields $
D,D^{*},f,f^{*},h,h^{*},g$ and $g^{*}$ play all the role of
auxiliary fields. The bosonic fields $s,s^{*},R_{\mu \nu },S_{\mu
\nu },\rho ,\rho ^{*},\phi ,\phi ^{*},r,r^{*}$ and the fermionic
fields $\xi ,\bar{\xi},\tau , \bar{\tau},F,\bar{F},\chi
,\bar{\chi},G,\bar{G},\zeta ,\bar{\zeta}$ work as background
fields also responsible for the breaking the Lorentz invariance.

\section{$N=4$-Lorentz-violating term}

In a very close analogy to the procedure adopted in the previous
section, we succeed in writing down the $N=4$ model by means of a
reduction from 10 to 4 dimensions. We propose for $D=10$ the
Chern-Simons term in the form
\begin{equation}
{\cal L}_{br}=\varepsilon ^{\hat{\mu}\hat{\nu}\hat{\kappa}\hat{\lambda}\hat{%
\rho}\hat{\sigma}\hat{\delta}\hat{\tau}\hat{\beta}\hat{\gamma}}A_{\hat{\mu}%
}\partial _{\hat{\nu}}A_{\hat{\kappa}}T_{\hat{\lambda}\hat{\rho}\hat{\sigma}%
\hat{\delta}\hat{\tau}\hat{\beta}\hat{\gamma}}.  \label{21}
\end{equation}
The background tensor $T_{\hat{\lambda}\hat{\rho}\hat{\sigma}}$
has 120 components, and we can redefine it as
\[
T_{\hat{\lambda}\hat{\rho}\hat{\sigma}}\equiv (R_{\rho \sigma
}^{I};\partial _{\mu }v;\partial _{\mu }u^{IJ}\,),
\]
where $\hat{\mu}=\mu ,4,5,6,7,8,9$ is the space-time index and $%
I,J=1,2,3,4,5,6$ is an internal index. We consider that there is
no dependence of the fields on the
$x^{4},x^{5},x^{6},x^{7},x^{8},x^{9}$ coordinates. Then, we have 6
anti-symmetric tensor fields $R_{\rho \sigma }^{I}$ with 6
components each one and 15 vectors written as gradients of 15
scalars represented by the anti-symmetric index $I,J.$ Therefore,
the number of independent components is reduced to 52.

Next, we need to redefine the gauge field as $A_{\hat{\mu}}\equiv
(A_{\mu };\varphi ^{I}\,,\,I=1,2,3,4,5,6)$ where $\varphi ^{I}$ is
real scalar
fields. Observing that $\varepsilon ^{\hat{\mu}\hat{\nu}\hat{\kappa}\hat{%
\lambda}\hat{\rho}\hat{\sigma}\hat{\delta}\hat{\tau}\hat{\beta}\hat{\gamma}%
}A_{\hat{\mu}}A_{\hat{\kappa}}\partial _{\hat{\nu}}T_{\hat{\lambda}\hat{\rho}%
\hat{\sigma}\hat{\delta}\hat{\tau}\hat{\beta}\hat{\gamma}}=0,$ we
obtain, integrating by parts, the Lagrangian as follows:
\begin{equation}
{\cal L}_{br}=-\frac{1}{4}\varepsilon ^{\mu \nu \kappa \lambda
}F_{\mu \nu }A_{\kappa }\partial _{\lambda
}v+\frac{1}{4}\varepsilon ^{\mu \nu \kappa \lambda }F_{\mu \nu
}\varphi ^{I}R_{\kappa \lambda }^{I}+\frac{1}{2}\varphi
^{I}\partial _{\nu }\varphi ^{J}\partial ^{\nu }u^{IJ}. \label{22}
\end{equation}
This is the bosonic sector of the action term to be
supersymmetrized. In this way, is necessary to define new fields
to be partners inside the superfields. They are similar to the
procedure of the previous section, but now there are internal
index. In terms of superfields, we have two sectors:

\begin{eqnarray*}
Gauge\,\, Sector &:&\,\,\,\,\,\{V,\Phi ^{I}\} \\
Background\,\, Sector \,\, &:&\,\,\,\,\,\{\Sigma
_{a}^{I},\bar{\Sigma}^{\dot{a}I},S,\bar{S},R^{IJ},\bar{R}^{IJ}\},
\end{eqnarray*}
and, in components these two sectors encompass the fields cast
below:
\begin{eqnarray*}
Bosonic\,\, gauge\,\, Sector &:&\,\,\,\,\,\{A_{\mu
},\varphi ^{I},\varphi ^{*I}\} \\
Fermionic\,\, gauge\,\, Sector &:&\,\,\,\{\lambda
,\bar{\lambda},\psi
^{I},\bar{\psi}^{I}\} \\
Bosonic\,\, background\,\, Sector &:&\,\,\,\,\,%
\{s,s^{*},R_{\mu \nu }^{I},\rho ^{I},\rho ^{*I},r^{IJ},r^{*IJ}\} \\
Fermionic\,\, background\,\, Sector &:&\,\,\,\,\,\{\xi ,\bar{\xi},\tau ^{I},%
\bar{\tau}^{I},F^{I},\bar{F}^{I},\zeta ^{IJ},\bar{\zeta}^{IJ}\}.
\end{eqnarray*}

Based on dimensional analysis arguments for the bosonic sector, as
it has been done for the $N=2$ case, and noticing that some
superfields now have internal symmetry index, we propose the
following $N=4$ supersymmetric action:

\begin{eqnarray}
{\cal S}_{br} &=&\int d^{4}xd^{2}\theta d^{2}\bar{\theta}[\frac{1}{4}%
W^{a}(D_{a}V)S+\frac{1}{4}\bar{W}_{\dot{a}}(\bar{D}^{\dot{a}}V)\bar{S}+\frac{%
i}{4}\delta (\bar{\theta})W^{a}(\Phi ^{I}+\bar{\Phi}^{I})\Sigma
_{a}^{I}
\label{40} \\
&&-\frac{i}{4}\delta (\theta )\bar{W}_{\dot{a}}(\Phi ^{I}+\bar{\Phi}^{I})%
\bar{\Sigma}^{\dot{a}I}+\frac{1}{4}\Phi ^{I}\bar{\Phi}^{J}(R^{IJ}-\bar{R}%
^{IJ})],  \nonumber
\end{eqnarray}

We can observe that the action (\ref{40}) is invariant under $N=1$%
-supersymmetry and there is a larger symmetry, the
$N=4$-supersymmetry as well.

This $N=4$ Lagrangian in its component-field version reads as
follows:
\begin{eqnarray}
{\cal L}_{br} &=&+\frac{i}{8}\partial _{\mu }(s-s^{*})\varepsilon
^{\mu \kappa \lambda \nu }F_{\kappa \lambda }A_{\nu
}-\frac{1}{8}(s+s^{*})F_{\mu
\nu }F^{\mu \nu }+D^{2}(s+s^{*})  \nonumber \\
&&-\frac{1}{2}is\lambda \sigma ^{\mu }\partial _{\mu }\bar{\lambda}-\frac{1}{%
2}is^{*}\bar{\lambda}\bar{\sigma}^{\mu }\partial _{\mu }\lambda -\frac{1}{2%
\sqrt{2}}\lambda \sigma ^{\mu \nu }F_{\mu \nu }\xi +\frac{1}{2\sqrt{2}}\bar{%
\lambda}\bar{\sigma}^{\mu \nu }F_{\mu \nu }\bar{\xi}  \nonumber \\
&&+\frac{1}{4}\lambda \lambda h+\frac{1}{4}\bar{\lambda}\bar{\lambda}h^{*}-%
\frac{1}{\sqrt{2}}\lambda \xi
D-\frac{1}{\sqrt{2}}\bar{\lambda}\bar{\xi}D
\nonumber \\
&&\frac{1}{16}\varepsilon ^{\mu \nu \kappa \lambda }F_{\mu \nu
}(\varphi
^{I}+\varphi ^{*I})(B_{\kappa \lambda }^{I}+B_{\kappa \lambda }^{*I})+\frac{i%
}{8}F^{\mu \nu }(B_{\mu \nu }^{I}-B_{\mu \nu }^{*I})(\varphi
^{I}+\varphi
^{*I})  \nonumber \\
&&-\frac{i\sqrt{2}}{8}\tau ^{I}\sigma ^{\mu \nu }\psi ^{I}F_{\mu \nu }-\frac{%
i\sqrt{2}}{8}\bar{\tau}^{I}\bar{\sigma}^{\mu \nu }\bar{\psi}^{I}F_{\mu \nu }+%
\frac{1}{4}\tau ^{I}\sigma ^{\mu }\partial _{\mu
}\bar{\lambda}(\varphi
^{I}+\varphi ^{*I})  \nonumber \\
&&-\frac{1}{4}\bar{\tau}^{I}\bar{\sigma}^{\mu }\partial _{\mu
}\lambda (\varphi ^{I}+\varphi ^{*I})+\frac{i\sqrt{2}}{4}\psi
^{I}\sigma ^{\mu \nu }B_{\mu \nu }^{I}\lambda
+\frac{i\sqrt{2}}{4}\bar{\psi}^{I}\bar{\sigma}^{\mu
\nu }B_{\mu \nu }^{*}\bar{\lambda}^{I}  \nonumber \\
&&-\frac{i}{2}D(\varphi ^{I}+\varphi ^{*I})\rho ^{I}+\frac{i}{2}%
D^{*}(\varphi ^{I}+\varphi ^{*I})\rho ^{*I}  \nonumber \\
&&+\frac{i\sqrt{2}}{8}\lambda \psi ^{I}\rho ^{I}-\frac{i\sqrt{2}}{8}\bar{%
\lambda}\bar{\psi}^{I}\rho ^{*I}-\frac{i\sqrt{2}}{4}D\psi ^{I}\tau ^{I}+%
\frac{i\sqrt{2}}{4}D^{*}\bar{\psi}^{I}\bar{\tau}^{I}  \label{30} \\
&&+\frac{i}{4}f^{I}\lambda \tau ^{I}-\frac{i}{4}f^{*I}\bar{\lambda}\bar{\tau}%
^{I}+\frac{i}{4}(\varphi ^{I}+\varphi ^{*I})\lambda F^{I}-\frac{i}{4}%
(\varphi ^{I}+\varphi ^{*I})\bar{\lambda}\bar{F}^{I}  \nonumber \\
&&+\frac{1}{8}\varphi ^{I}\partial _{\mu }\varphi ^{*J}\partial
^{\mu }(r^{IJ}+r^{*IJ})-\frac{1}{8}\varphi ^{*J}\partial _{\mu
}\varphi
^{I}\partial ^{\mu }(r^{IJ}+r^{*IJ})  \nonumber \\
&&+\frac{1}{4}\partial ^{\mu }\varphi ^{I}\partial _{\mu }\varphi
^{*J}(r^{IJ}-r^{*IJ})-\frac{1}{8}\varphi ^{I}\varphi ^{*J}\square
(r^{IJ}-r^{*IJ})-\frac{i}{4}\psi ^{I}\sigma ^{\mu }\partial _{\mu }\bar{\psi}%
^{J}(r^{IJ}-r^{*IJ})  \nonumber \\
&&+\frac{1}{4}f^{I}f^{*J}(r^{IJ}-r^{*IJ})+\frac{i}{4}\psi ^{I}\sigma ^{\mu }%
\bar{\psi}^{J}\partial _{\mu }r^{*IJ}  \nonumber \\
&&-\frac{i}{4}\varphi ^{I}\zeta ^{IJ}\sigma ^{\mu }\partial _{\mu }\bar{\psi}%
^{J}+\frac{i}{4}\varphi ^{*J}\psi ^{I}\sigma ^{\mu }\partial _{\mu }\bar{%
\zeta}^{IJ}+\frac{i}{4}\psi ^{I}\sigma ^{\mu
}\bar{\zeta}^{IJ}\partial _{\mu
}\varphi ^{*J}  \nonumber \\
&&+\frac{1}{4}\varphi ^{I}f^{*J}g^{IJ}-\frac{1}{4}f^{I}\varphi ^{*J}g^{*IJ}-%
\frac{1}{4}f^{*J}\psi ^{I}\zeta ^{IJ}+\frac{1}{4}f^{I}\bar{\psi}^{J}\bar{%
\zeta}^{IJ}.  \nonumber
\end{eqnarray}
We can ascertain the presence of the bosonic sector (\ref{22}) by
means of the terms below:
\begin{eqnarray*}
\frac{i}{8}\partial _{\mu }(s-s^{*})\varepsilon ^{\mu \kappa
\lambda \nu }F_{\kappa \lambda }A_{\nu }
&=&-\frac{1}{4}\varepsilon ^{\mu \nu \kappa
\lambda }F_{\mu \nu }A_{\kappa }\partial _{\lambda }v, \\
\frac{1}{16}\varepsilon ^{\mu \nu \kappa \lambda }F_{\mu \nu
}(\varphi
^{I}+\varphi ^{*I})(B_{\kappa \lambda }^{I}+B_{\kappa \lambda }^{*I}) &=&%
\frac{1}{4}\varepsilon ^{\mu \nu \kappa \lambda }F_{\mu \nu
}\varphi
^{I}R_{\kappa \lambda }^{I}, \\
\frac{1}{8}\varphi ^{I}\partial _{\mu }\varphi ^{*J}\partial ^{\mu
}(r^{IJ}+r^{*IJ})-\frac{1}{8}\varphi ^{*J}\partial _{\mu }\varphi
^{I}\partial ^{\mu }(r^{IJ}+r^{*IJ}) &=&\frac{1}{2}(\varphi
^{I}\partial _{\nu }\varphi ^{J}+\beta ^{I}\partial _{\nu }\beta
^{J})\partial ^{\nu }u^{IJ}.
\end{eqnarray*}
We can notice that this Lagrangian fairly accommodates the $N=4$
bosonic sector (\ref{22}). We re-obtain here the $N=1$ and $N=2$
supersymmetrization of the Chern-Simons term presented in
ref.\cite{3} and in (\ref{14}), respectively. We notice that $N=4$
Lagrangian is similar to $N=2$ but now existing an internal index
in same fields. The fields $\beta ^{I},\,t,\,\,u^{IJ}$ and $\,\rho
^{IJ},$ that do not appear in the bosonic Lagrangian (\ref{22}),
were introduced in order to keep the bosonic and fermionic degrees
of freedom in equal number. We can see that the bosonic fields
$D,D^{*},f^{I},f^{*I},h,h^{*},g^{IJ}$ and $g^{*IJ}$ works as
auxiliary fields. The bosonic fields $s,s^{*},R_{\mu \nu
}^{I},\rho
^{I},\rho ^{*I},r^{IJ},r^{*IJ}$ and the fermionic fields $\xi ,\bar{\xi}%
,\tau ^{I},\bar{\tau}^{I},F^{I},\bar{F}^{I},\zeta
^{IJ},\bar{\zeta}^{IJ}$ work as background fields breaking the
Lorentz invariance.

\section{Concluding remarks and comments}

In the important context of studying the gauge invariant
Lorentz-violating term formulated as a Chern-Simons action term ,
we propose here its $N=2$ and $N=4$ supersymmetric versions. This
program can be carry out in a simple way with the help of a
dimensional reduction method; here, we have chosen the method
\`{a} la Scherk, but it would also be interesting to contemplate
other possibilities, such  as the procedures  \`{a} la Legendre or
\`{a} la Kaluza-Klein. With our reduction scheme,
we could treat the extended supersymmetric version in terms of simple $%
N=1 $ superspace to supersymmetrize the Chern-Simons like term, as
proposed by Jackiw, written in terms of a constant background
vector here parametrized as the
gradient of the scalar function $\alpha +\beta _{\mu }x^{\mu },$ where $%
\alpha $ and $\beta ^{\mu }$ are constants.

Another interesting point we should consider is the possibility,
once we have now the full set of SUSY partners of the
Lorentz-breaking vector, to express the central charges of the
extended models whenever topologically non-trivial configurations
are taken into account. This would allow us to impose bounds on
the central charges in terms of the phenomenological constraints
already imposed on the vector responsible for the Lorentz
covariance breakdown.


\end{document}